\documentclass[twocolumn]{aastex631}

\usepackage{graphicx,epstopdf}
\usepackage[T1]{fontenc}
\usepackage[utf8]{inputenc}
\usepackage{hyperref}
\usepackage{color}
\usepackage{subfigure}
\usepackage{amsmath}
\usepackage{amssymb}
\usepackage{epsfig}
\usepackage{amsfonts}
\usepackage{mathtools}
\usepackage{bbm}
\usepackage{bm}
\usepackage{float}
\usepackage{xcolor}

\def\rvec{{\bf r}}

\def\kvec{{\bf k}}
\def\pvec{{\bf p}}

\def\bra#1{\left\langle#1\right|}
\def\ket#1{\left|#1\right\rangle}

\def\he#1{$^{#1}$He}

\def\KF{k_{\mathrm F}}

\def\S{{\bf S}}

\def\1{\mathbbm{1}}

\def\bra#1{\bigl\langle{ #1} \bigr|}
\def\ket#1{\bigl|{ #1} \bigr\rangle}

\def\bsigma{{\bm\sigma}}

\def\he#1{$^{#1}$He}

\shorttitle{triplet pairing }
\shortauthors{Krotscheck et al.}
\begin{document}

\title{Triplet pairing in neutron matter}

\email{eckhardk@buffalo.edu,ppapakon@ibs.re.kr,jwang97@buffalo.edu}

\author[0000-0001-7739-1255]{Eckhard Krotscheck}
\affiliation{Department of Physics,
  University at Buffalo, SUNY,
  Buffalo NY 14260, USA\\
  and\\
Institut f\"ur Theoretische Physik,
Johannes Kepler Universit\"at,
A 4040 Linz, Austria}

\author[0000-0002-7119-6667]{Panagiota Papakonstantinou}
\affiliation{Rare Isotope Science
  Project, Institute for Basic Science, Daejeon 34000, Korea}

\author[0000-0002-4667-3791]{Jiawei Wang}
\affiliation{Department of Physics,
  University at Buffalo, SUNY,
  Buffalo NY 14260, USA
}

\begin{abstract}

The presence of superfluidity in neutron star interiors can affect the
cooling of neutron stars in intricate ways, enhancing certain
mechanisms and suppressing others. Model calculations employing
realistic nuclear potentials in Bardeen-Cooper-Schrieffer theory
generally suggest the development of a $^3$P$_2$-$^3$F$_2$ pairing
gap, and therefore the presence of superfluidity in dense neutron star
matter.  Improved models that go beyond conventional mean-field
calculations by including polarization effects suggest a suppression
of the triplet gap. We have evaluated the pairing interaction by
summing the ``parquet'' Feynman diagrams which include both ladder and
ring diagrams systematically, {\em plus\/} a set of important
non-parquet diagrams, making this the most comprehensive diagram-based
approach presently available.  Our results suggest a radical
suppression of the $^3$P$_2$-$^3$F$_2$ triplet pairing gap and an
enhancement of $^3$P$_0$ pairing.
\end{abstract}


\section{Introduction}
\label{sec:intro}

A prevailing and persistent question in the interpretation of cooling
data in neutron stars concerns the presence or
absence of superfluidity in the star's core. The most relevant kind of
observation involves the cooling curves of
neutron stars, {\em i.e.}, the evolution of surface temperature with
time \citep{PageReddy2006,StellarSuperfluids,YaH2003}.  Superfluidity
can induce competing effects: On one hand, it suppresses the direct
URCA process. On the other hand, it enhances the pair breaking and
formation mechanism when the temperature approaches the critical
temperature \citep{StellarSuperfluids}. More specifically, the
comparison of the measured luminosity and
inferred temperatures with theoretical predictions reveals the
neutrino emissivity, which in turn is affected by superfluidity..

The impact of pair breaking and formation on neutrino
  pair emissivity was first predicted by \cite{Flowers1976}, who
  examined the singlet channel. Later calculations showed that
  collective effects can suppress the process
  \cite{Leinson2001,Leinson2006,Sedrakian2007,Kolomeitsev2008,Steiner2009,Page2009}.
  Triplet pairing considered along or perpendicular to the
  quantization axis can lead to a different dependence of the neutrino
  energy loss on the gap \cite{Gusakov2002} compared to the singlet
  case. The suppression due to collective effects has been thought to
  be less dramatic in the triplet case, especially in the axial vector
  channel, although the exact value of the suppression coefficient is
  debatable \cite{Leinson2010, Shternin2011}. The question of triplet
  pairing development in the neutron star core is especially relevant
  in the description of the Cassiopeia A remnant's cooling. On one
  hand, one may have to assume some suppression of URCA processes,
  otherwise the object would be too cold. On the other hand, a
  speed-up mechanism is required to explain the later rapid
  cooling. One such mechanism could be Cooper pair breaking, which was
  explored in Refs. \cite{Shternin2011, Page2011, Shternin2022}. It is,
  however, not the only possible mechanism \cite{Yang2011}. Hybrid
  scenarios have also been considered \cite{Yang2011}.

Generally, it turns out that some observations are compatible with the
so-called ``Minimal Cooling Paradigm'', without drastically enhanced
neutrino emission processes, while others show enhanced emissivity, as
discussed in Ref.~\citet{StellarSuperfluids}. This seems to indicate a
critical interplay of competing factors in determining the cooling
curve of a neutron star.  More recently, an analysis of thermal states
of neutron stars in soft X-ray transients suggests a low critical
temperature and small gap for the triplet superfluid in the core
region \citep{HaS2017}. As mentioned above, the analysis of the
cooling neutron star in Cassiopeia A, on the other hand, supports the
presence of triplet superfluidity \citep{Sthernin2022}.  We note that
the proton fraction, which depends on the theoretically calculated
equation of state and is somewhat uncertain, and affects the emissivity,
too.  An analysis of the accreting neutron star MXB 1659-29 suggests
that superfluidity is not needed to describe its luminosity, as long
as the symmetry-energy slope does not exceed 80~MeV (which is indeed
considered too high according to most existing constraints
\citep{RevModPhys.89.015007}) \citep{MFC2022}.  The empirical evidence
for the development of triplet superfluidity in neutron stars is
therefore rather inconclusive. It seems therefore very important to
tackle the question as well as possible from the theoretical,
microscopic point of view in order to make real progress towards
interpreting the observational data.

The superfluid phase transition in neutron matter has been studied
microscopically for decades; see, for example,
Refs. \citet{SedrakianClarkBCSReview} and \citet{SchuckBCS2018} for
recent reviews and extensive compilations of the relevant literature.
We can basically distinguish four types of calculations:
\begin{enumerate}
\item{} Calculation of the superfluid gap at the mean-field level
  using bare, more or less realistic nuclear interactions. See
    for example Refs. \citet{Baldo3P23F2,KKC96}.
\item{} It was already realized very early that
  medium polarization effects can have a profound influence on the
  superfluid transition temperature
  \citep{CKY76,JWCgap,Wam93,SchulzePLB96}.  The presently most
  sophisticated treatment of these effects is found in Ref.
  \citet{SchwenkFriman2004} also with the
  conclusion that the effect of polarization can be quite dramatic. To
  calculate polarization corrections, assumptions on the quasiparticle
  interaction must be made.
\item{} The inclusion of many-body effects may be
  traced to a formulation of correlated basis function theory
  \citep{FeenbergBook} for superfluid systems
  \citep{HNCBCS,Fantonipairing} or extensions of
  Brueckner-Bethe-Goldstone theory
  \citep{SPR2001,PhysRevC.94.025802,JLTP189_234}. Generally, the
  method can be mapped onto a regular BCS-like theory with effective
  interactions, the essential task then being the calculation of these
  effective interactions with a trustworthy accuracy. When executed to
  a sufficiently high level, microscopic calculations also provide the
  quasiparticle interaction needed for the calculation of polarization
  corrections.
\item{} Some Monte Carlo calculations exist for S-wave pairing in
  low-density neutron matter \citep{GC2008}. Our results of
  Ref. \citet{v3bcs} agree quite well with these calculations;
  whether the extension of Monte Carlo methods for systems where
  complicated tensor- and spin-orbit components of the interaction are
  essential remains to be seen.
\end{enumerate}
In the above classification of methods, our work belongs to the third
category.  We calculate the pairing interaction by a self-consistent
summation of all ring- and ladder-diagrams which corresponds, in the
language of Jastrow-Feenberg theory, to the optimized
Fermi-Hypernetted-Chain (FHNC-EL) summation method. We also include
important non-parquet diagrams which correspond in Jastrow-Feenberg
theory, to the so-called ``commutator corrections''. The relationship
between diagrammatic perturbation theory \citep{BaymKad} and FHNC-EL
has been established rigorously for bosons \citep{parquet1,parquet2}
and for the most relevant diagrammatic substructures for fermions
\citep{fullbcs}.  As we will see, the polarization effects introduced
in this manner lead to the suppression of the triplet
$^3$P$_2$-$^3$F$_2$ pairing, in agreement with previous studies
\citep{SchwenkFriman2004}, and at the same time to
  an enhancement of the $^3P_0$ pairing gap. More precisely, the
effect comes from the suppression of the spin-orbit potential due to
short-ranged screening \citep{v4}.
  
\section{Microscopic Theory}

Common to all treatments of superfluidity in nuclear and
neutron(star) matter is the equation for the pairing gap
$\Delta(\kvec)$,
\begin{equation}
  \Delta(\kvec) = -\frac{1}{2}\sum_{\kvec'}\bra{\kvec}V\ket{\kvec'}
  \frac{\Delta(\kvec')}{E(\kvec')}
  \label{eq:GeneralGap}
\end{equation}
where
\begin{equation}
  E(\kvec) = \sqrt{(\epsilon(k)-\mu)^2 + \left|\Delta(\kvec)\right|^2}
\end{equation}
Above, the $\epsilon(k)$ are the single-particle energies
  of the normal system, and $\mu$ is the chemical potential.
Typically, Eq. \eqref{eq:GeneralGap} is decoupled in different angular
momentum channels by approximating the quasiparticle energy denominator
$E(\kvec)$ by its angle average \citep{Baldo3P23F2}.
\begin{equation}
  \begin{split}
  E(\kvec) &\approx \sqrt{(\epsilon(k)-\mu)^2 + D^2(k)}\,,\\
  D^2(k) &= \frac{1}{4\pi}\int d\Omega_\kvec \left|\Delta(\kvec)\right|^2\,.
  \label{eq:angav}
  \end{split}
\end{equation}
This is justified by the finding that the five solutions of the
$^3$P$_2$ gap equation are nearly degenerate
\citep{10.1143/PTP.46.114}. The gap function $\Delta(\kvec)$ can be
expanded in spherical harmonics, In general, the gap equation
couples different angular momenta and becomes a matrix equation of the
form
\begin{equation}
  \Delta^{(\ell)}(k) =\label{eq:multigap}
  -\frac{1}{2}\sum_{\ell'}\int \frac{d^3k'}{(2\pi)^3}
  \frac{V_{\ell\,\ell'}(k,k')\Delta^{(\ell')}(k')}{
\sqrt{(\epsilon(k')-\mu)^2 + D^2(k')}}\,.
\end{equation}
 
Without going into the technical details, let us summarize what it
takes to formulate a microscopic theory of superfluidity in a system
where many-body effects are expected to be relevant:

\begin{enumerate}
  \item{} The interaction between individual nucleons is complicated
    and often formulated in individual angular momentum and spin
    channels \citep{Reid68,Reid93}.  For our present purpose of
    summing vast arrays of Feynman diagrams, the representation in an
    operator basis \citep{Day81,AV18} is much more practical and
    therefore preferred:
\begin{equation}
\hat v (i,j) = \sum_{\alpha=1}^n v_\alpha(r_{ij})\,\hat O_\alpha(i,j)\,.
\label{eq:vop}
\end{equation}
Here $r_{ij}=\left|\rvec_i-\rvec_j\right|$ is the distance between
particles $i$ and $j$, and the $\hat O_\alpha(i,j)$ are operators
acting on the spin, isospin, and possibly the relative angular
momentum variables of the individual particles.  According to the
number of operators $n$, the potential model is referred to as a $v_n$
model potential. Semi-realistic models for nuclear matter keep at
least the six to eight base operators
\begin{eqnarray}
\hat O_1(i,j;\hat\rvec_{ij})
        &\equiv& \hat O_c = \1\,,
\nonumber\\
\hat O_3(i,j;\hat\rvec_{ij})
        &\equiv& \hat O_\sigma(i,j;\hat\rvec_{ij}) = {\bsigma}_i \cdot {\bsigma}_j\,,
\nonumber\\
\hat O_5(i,j;\hat\rvec_{ij})&\equiv&\hat O_{\rm S}(i,j;\hat\rvec_{ij})
= S_{ij}(\hat\rvec_{ij})\nonumber\\
      &\equiv& 3({\bm\sigma}_i\cdot \hat\rvec_{ij})
      ({\bsigma}_j\cdot \hat\rvec_{ij})-{\bsigma}_i \cdot {\bsigma}_j\,,
      \nonumber\\
  \hat O_7(i,j;\rvec_{ij},\pvec_{ij})
  &\equiv&\hat O_{\rm LS}(i,j;\hat\rvec_{ij})= \rvec_{ij}\times\pvec_{ij}\cdot\S\,,
  \nonumber\\
      \hat O_{2\alpha}(i,j;\hat\rvec_{ij}) &=& \hat O_{2\alpha-1}(i,j;\hat\rvec_{ij})
      {\bm\tau}_i\cdot{\bm\tau}_j\,,
  \label{eq:operator_v8}
\end{eqnarray}
where $\S\equiv\frac{1}{2}(\bsigma_i+\bsigma_j)$ is the total spin,
and $\pvec_{ij}=\frac{1}{2}(\pvec_i-\pvec_j)$ is the relative momentum
operator of the pair of particles.  In neutron matter, the operators
are projected to the isospin=1 channel.

\begin{figure}[H]
  \centerline{\includegraphics[width=0.7\columnwidth]{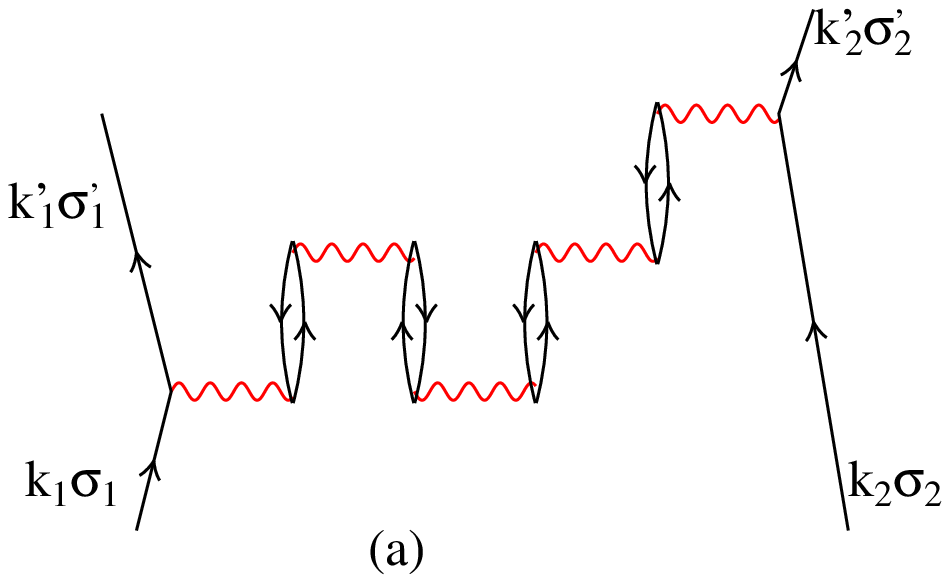}}
  \centerline{
    \includegraphics[width=0.35\columnwidth]{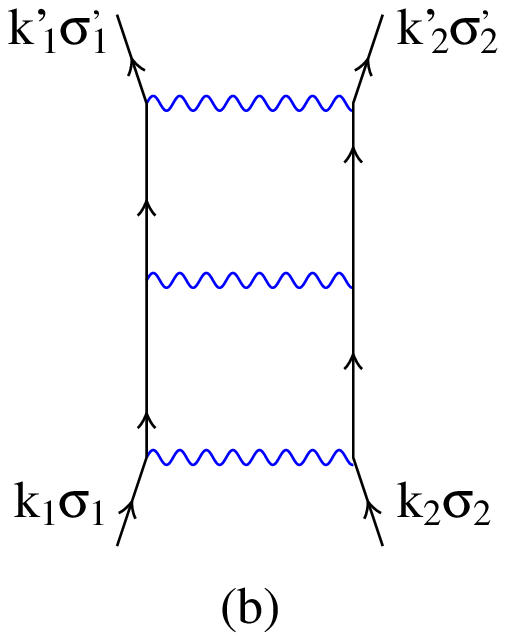}
    \includegraphics[width=0.35\columnwidth]{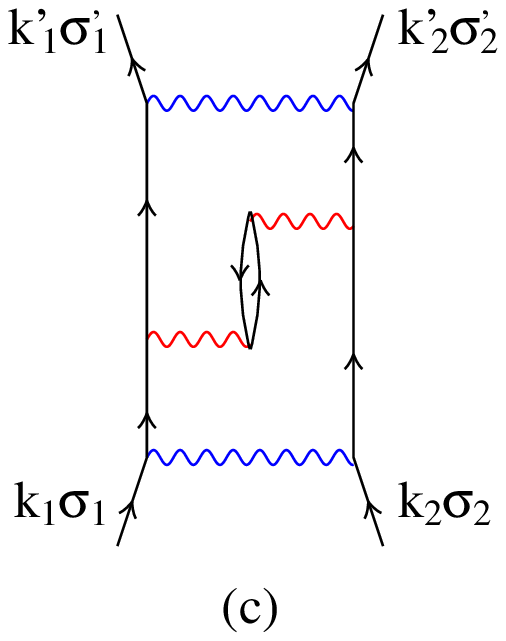}}
  \centerline{
    \includegraphics[width=0.35\columnwidth]{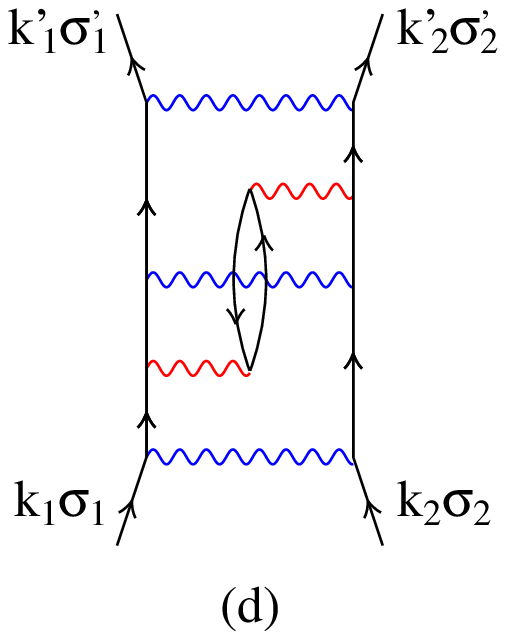}
    \includegraphics[width=0.35\columnwidth]{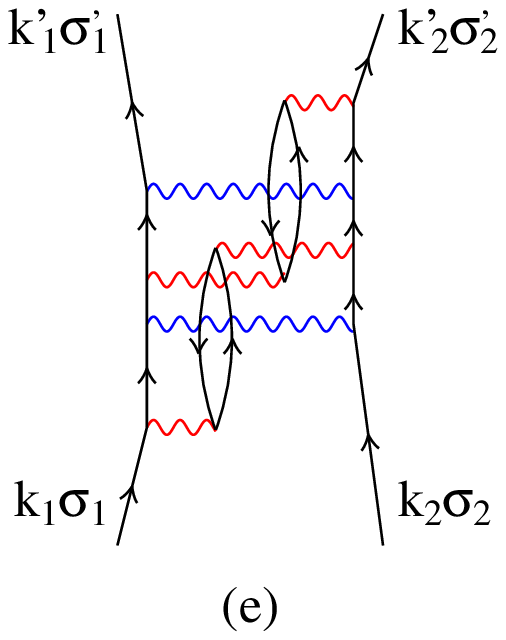}}
  
     \caption{(color online) Examples of the Feynman diagrams
       included in the effective interaction.  Each red wavy line
       represents a particle-hole irreducible vertex.  Each blue
       wavy line depicts a bare interaction.  (a)
       is a simple example of a ``chain''
       diagram.  (b) and (c) are simple example of ``ladder''
       diagrams which may just consist of bare interactions (diagram
       (b)) or the rungs of the ladder may contain chain (diagrams
       (c)). Diagrams (d) and (e) show simple examples of ``twisted
       chain'' diagrams.
    \label{fig:diagrams} 
   } 
\end{figure}

\item{} The inclusion of ring diagrams
  (Figure \ref{fig:diagrams}a) describes the above-mentioned
  polarization effects. The driving interaction, depicted as a wavy
  line in fig. \ref{fig:diagrams}, is related to the quasiparticle
  interaction. It is important to realize that this quantity must not
  be identified with the $G$-matrix of Brueckner-Bethe-Goldstone
  theory. This is most easily seen in self-bound systems like nuclear
  matter or quantum fluids: The $G$ matrix determines the binding
  energy and is, hence, an {\em attractive\/} interaction whereas the
  quasiparticle interaction is related to the incompressibility and
  is, therefore, {\em repulsive\/}.

\item{} The inclusion of ladder diagrams describes the {\em
  short-ranged\/} structure of the system. 
  In the simplest case, the driving interaction is just the bare interaction;
  this defines the $G$ matrix. The rungs of
  these ladders can also include chain diagrams.

\item{} ``Twisted chain'' diagrams. These do {\em not\/} belong to the
  ``parquet'' class, but they can be important in cases where the bare
  interaction is very different in the spin-singlet and the
  spin-triplet channels \citep{SpinTwist,v3twist} which is the case
  for modern nucleon-nucleon interactions.

\end{enumerate}

Examples of ladder and twisted-chain diagrams are also shown in
fig. \ref{fig:diagrams}.

\section{Results}

For our calculations we have used the $v_8$ form
of the nucleon-nucleon interaction for the Reid interaction
\citep{Reid68} as formulated in Ref. \citet{Day81} as well as the
Argonne potential \citep{AV18} as represented in
eqs. \eqref{eq:vop}-\eqref{eq:operator_v8}.  We have calculated the
pairing interaction as described in Ref. \citet{v3bcs}. In addition,
we have included the spin-orbit term which is necessary for triplet
pairing but plays no role for pairing in the $^1$S$_0$ channel. As
observed in a series of recent publications
\citep{v3eos,v3bcs,v3twist,v4}, the results for the Reid and Argonne
interactions are very close. Accordingly, we
report here only the ones obtained with the Reid
interaction.

  Throughout our calculations, we have utilized a
  non-interacting single-particle spectrum $\epsilon(k) = \hbar^2
  k^2/2m$. One can go beyond such a simplifying approximation by
  either using the single-particle spectrum predicted by correlated
  basis functions theory \cite{HNCBCS} or improve upon that by
  including dynamic effects.  These methods have been very successful
  for explaining the physical mechanisms leading to the strong
  effective mass enhancement in $^3$He \cite{he3mass}, but they
  appeared here to be a technical overkill. The effective mass ratio
  is expected to be between 1.05 and 0.95 \cite{ectpaper} and
  basically scales the gap whereas we will see that the effect we are
  reporting can change the gap by two orders of magnitude.

Given a bare interaction of the form \eqref{eq:vop},
  the effective interaction entering the gap equation
  \eqref{eq:GeneralGap} has {\em a priori\/} the same operator
  form. The individual angular momentum components are generated from
  these, in the $\ket{k, \ell, j}$ basis
\begin{subequations}
  \label{eq:veffs}
\begin{align}
    V_{^3P_0-^3P_0}(k,k') &= \bra{k,1,0} v^{\rm eff}_{c} - 4 v^{\rm eff}_{\rm S}
    -2 v^{\rm eff}_{\rm LS}\ket{k',1,0}\,,\\
    V_{^3\mathrm{P}_2-^3\mathrm{P}_2}(k,k') &=
    \bra{k,1,2}v^{\rm eff}_{c} - \frac{2}{5} v^{\rm eff}_{\rm S}
      + v^{\rm eff}_{\rm LS}\ket{k',1,2}\,,\\
    V_{^3\mathrm{P}_2-^3\mathrm{F}_2}(k,k') &=
    \frac{6}{5}\sqrt{6}\bra{k,1,2} v^{\rm eff}_{\rm S}\ket{k',3,2}\,,
    \\
    V_{^3\mathrm{F}_2-^3\mathrm{F}_2}(k,k') &=
     \bra{k,3,2}v^{\rm eff}_{c} - \frac{8}{5} v^{\rm eff}_{\rm S}
      -4 v^{\rm eff}_{\rm LS}\ket{k,3,2}\,.
  \end{align}
\end{subequations}
where the three effective interactions $v^{\rm eff}_{c}$, $v^{\rm eff}_{\rm S}$
and $v^{\rm eff}_{\rm LS}$ are projections of the general operator
structure \eqref{eq:vop} to spin and isospin 1, they are normally non-local operators in momentum
space \cite{v3bcs,v4}.

\begin{figure}
  \centerline{\includegraphics[width=0.6\columnwidth,angle=-90]{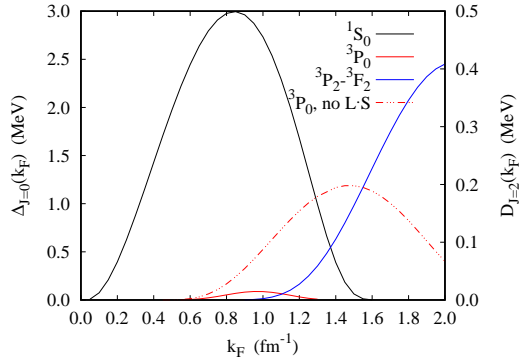}}
  \caption{(color online) The figure shows dependence of the superfluid gap for
    $^1$S$_0$ (black solid line), $^3$P$_2$-$^3$F$_2$ (blue solid line)
    and $^3$P$_0$ pairing (red solid line) with the Reid
    interaction on the Fermi wave number $\KF$. Also shown is the $^3$P$_0$ pairing gap when the
    spin-orbit interaction is turned off (red dash-dotted line). Note
    that the scale on the left-hand side of the figure refers to
    $^1$S$_0$ and  $^3$P$_0$ states, whereas the scale on the right
    refers to the  $^3$P$_2$-$^3$F$_2$ channel.
    \label{fig:baregaps}}
\end{figure}

The special role of the spin-orbit interaction has been pointed out in
Ref. \citet{Gezerlis2014}: "without an attractive spin-orbit
interaction, neutrons would form a $^3$P$_0$ superfluid, in which the
spin and orbital angular momenta are anti-aligned, rather than the
$^3$P$_2$ state, in which they are aligned." This statement is, of
course, based on properties of the bare interaction. We have, therefore
and as a first step calculated the pairing gap for the bare
interaction with and without the spin-orbit force; the results are
shown in fig. \ref{fig:baregaps}. Our results agree with those of
earlier work \citep{KKC96,Baldo3P23F2,KhodelClark2001}. As stated above
the results for the Argonne and Reid interactions are almost
identical (see also \citep{v3bcs}). We did not find pairing in the
$^3$P$_2$ channel.

We have also done the same calculation with the spin-orbit interaction
turned off. As expected \citep{Gezerlis2014}, the pairing in
$^3$P$_2$-$^3$F$_2$ states disappears, whereas there is a measurable $^3$P$_0$
gap.

\begin{figure}
  \centerline{\includegraphics[width=0.5\columnwidth,angle=-90]{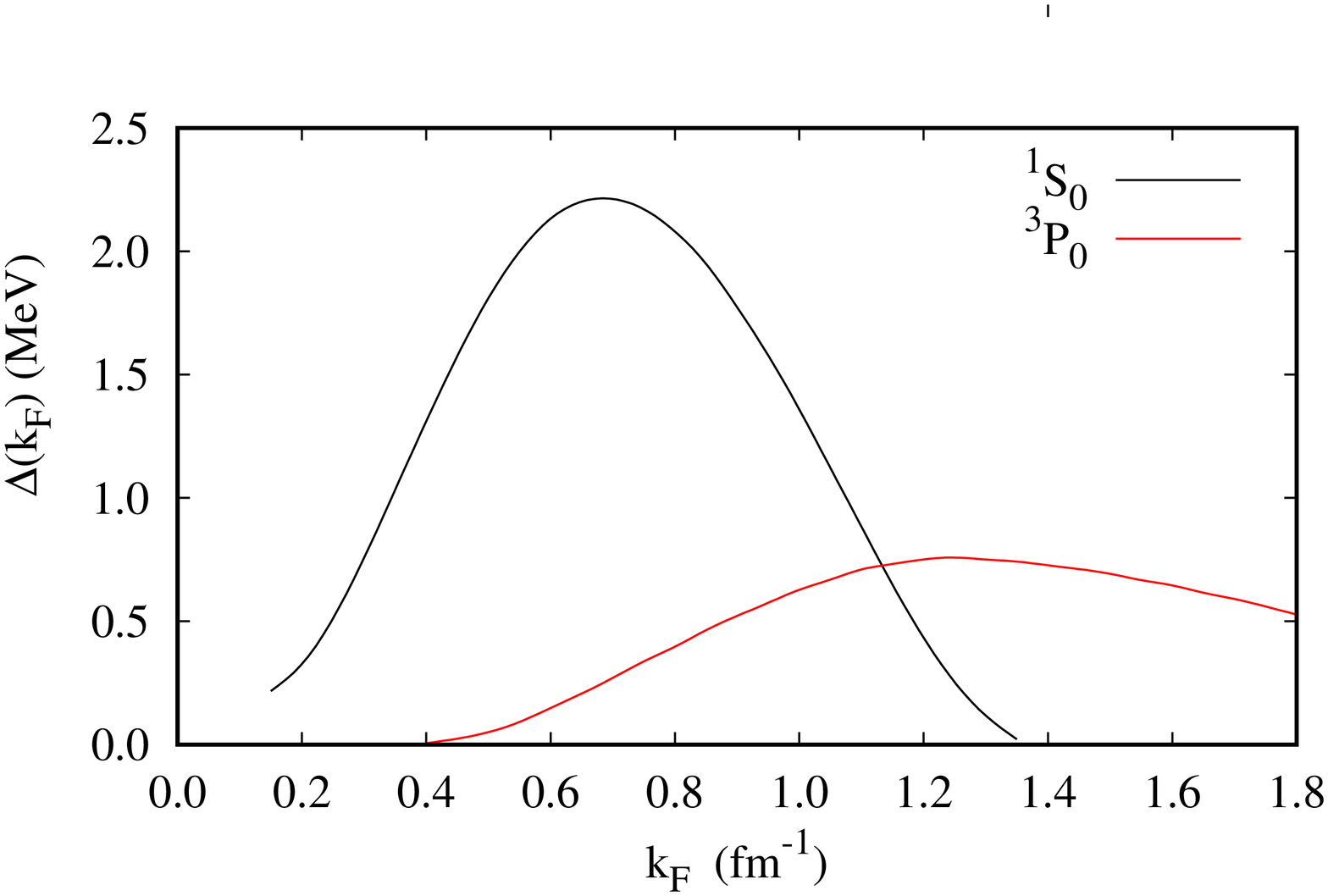}}
  \caption{(color online) The figure shows the superfluid gap for
    $^1$S$_0$ (black line), and the $^3$P$_0$ pairing (red line) for
    the effective interaction derived by parquet-diagram summation
    from the Reid interaction.
    \label{fig:reidgaps}}
\end{figure}

When all the many-body effects described above are taken into account,
the situation changes drastically for P-wave pairing. Figure
\ref{fig:reidgaps} shows our results for S- and P-wave pairing
employing an effective interaction that includes all the parquet and
important ``beyond-parquet'' Feynman diagrams. S-wave pairing is
somewhat reduced; this is consistent with much of the earlier work and
simply due to the fact that the effect occurs at relatively low
density.  On the other hand, the two channels of $P$-wave pairing are
reversed; 
In particular, $^3$P$_2$-$^3$F$_2$ pairing becomes minute whereas we have a measurable $^3$P$_0$ gap.

\begin{figure}
  \centerline{\includegraphics[width=0.6\columnwidth,angle=-90]{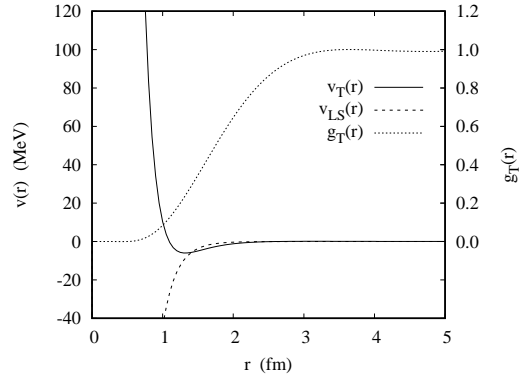}}
  \caption{The figure shows the interaction of a pair
    of neutrons in a spin-triplet state (solid line), the resulting
    spin-triplet pair distribution function $g_T(r)$ (short-dashed line)
and the spin-orbit potential (long-dashed line). \label{fig:V_and_g}}
\end{figure}

While the effect we are pointing out here seems to be
  an intricate consequence of high-level microscopic many-body theory,
  it is actually quite plausible as soon as one goes beyond mean-field
  pictures: The reason for nuclear saturation is the strong
  short-ranged repulsion that keeps the nucleons apart from each
  other. The effect is manifested in the {\em pair distribution
    function\/} $g(r)$ which is a normalized probability to find a
  pair of particles at a certain distance. Fig. \ref{fig:V_and_g}
  shows the function $g_T(r)$ for a pair of neutrons with parallel
  spin along with the bare central and spin-prbit
  interaction. Evidently, short-ranged screening has the effect that
  two neutrons never come close enough to each other so that they can
  ``see'' the attractive spin-orbit force. Thus, high-level many-body
  theory as utilized here is necessary for the {\em quantitative\/}
  determination of the screening effect, the effect is {\em
    qualitatively\/} quite plausible.

With that we have challenged an over 50 years old
narrative initiated by the pioneering work of Tamagaki {\em et al.\/}
\citep{Tamagaki70,10.1143/PTP.48.1517} that $^3$P$_2$ and
$^3$P$_2$-$^3$F$_2$ pairing prevails in neutron star matter. The
reason for this reversal is our result \citep{v4} that the spin-orbit
potential is strongly suppressed by many-body correlations.

We have limited our calculations to a density corresponding to $\KF
\le 1.8\,\mathrm{fm}^{-1}$.  At higher densities one expects that,
similar to \he3 \citep{polish}, more complicated many-body
correlations become important; the parquet summation technique for
operator-dependent interactions in these complicated cases still need
to be developed.

To summarize, we have shown that the most advanced microscopic
many-body theories predict the suppression of the in-medium spin-orbit
interaction, which leads unequivocally to the suppression of triplet
$^3$P$_2$-$^3$F$_2$ pairing in dense neutron matter.  Given that the
astronomical evidence for pairing in neutron star cores is already
inconclusive, our results clear the way to consider with more
confidence cooling scenarios which do not involve $^3$P$_2$
superfluidity.  At the same time, $^3$P$_0$ pairing cannot be
neglected and its effect on the cooling curve is worth the study.

\section*{Acknowledgments}
%\begin{acknowledgments}
  This work was supported, in part, by the College of Arts and
  Sciences of the University at Buffalo, SUNY (to J.W.). P.P. was
  supported by the Rare Isotope Science Project of the Institute for
  Basic Science funded by the Ministry of Science, ICT and Future
  Planning and the National Research Foundation (NRF) of Korea
  (2013M7A1A1075764).
%\end{acknowledgments}

\bibliography{papers}{} \bibliographystyle{aasjournal}

\end{document}